\documentclass{aa}
\usepackage{graphicx}

.tfm    
\def\lsim{\lower.5ex\hbox{$\; \buildrel < \over \sim \;$}}
\def\gsim{\lower.5ex\hbox{$\; \buildrel > \over \sim \;$}}
\def\be{\begin{equation}}
\def\ee{\end{equation}}
\def\eng{{\cal E}}
\def\vel{\vartheta}

\def\la{\lambda}

\begin{document}
\title{Computation of outflow rates from accretion disks around black holes}
\author{Santabrata Das
 \and Indranil Chattopadhyay
  \and A. Nandi
   \and Sandip K. Chakrabarti\thanks{\emph{Honorary Scientist,}
Centre for Space Physics, 114/v/1A Raja S.C. Mullick Rd., Kolkata 700047}}

\offprints {S. Das, \email{sbdas@bose.res.in}}

\institute{S. N. Bose National Centre for Basic Sciences, 
Salt Lake, Kolkata 700098, India}
%

\date{Received  15th May 2001  / Accepted 30th August 2001; Astronomy and Astrophysics, 379, 683, 2001}


\abstract{
We self-consistently estimate the outflow rate from the 
accretion rates of an accretion disk around a black hole 
in which both the Keplerian and the sub-Keplerian matter 
flows simultaneously. While Keplerian matter supplies 
soft-photons, hot sub-Keplerian matter supplies thermal 
electrons. The temperature of the hot electrons is decided by
the degree of inverse Comptonization of the soft photons. 
If we consider only thermally-driven flows from 
the centrifugal pressure-supported  boundary layer around
a black hole, we find that when the thermal electrons are 
cooled down, either because of the absence of the boundary 
layer (low compression ratio), or when the surface of the 
boundary layer is formed very far away, the outflow rate is 
negligible. For an intermediate size of this boundary layer
the outflow rate is maximal. Since the temperature of the 
thermal electrons also decides the spectral state of 
a black hole, we predict that the outflow rate should be
directly related to the spectral state.
}


\maketitle

\medskip

\section{Introduction}

Most of the galactic black hole candidates are known to undergo
spectral state transitions (Tanaka \& Lewin, 1995; Chakrabarti \& 
Titarchuk, 1995, hereafter CT95; Ebisawa et al. 1996). Two common states are 
the so-called hard state and the soft state. In the former, soft-X-ray 
luminosity is low and the energy spectral index $\alpha \sim 0.5$ ($E_\nu 
\propto \nu^{-\alpha}$) in the 2-10keV range. In the latter state, 
the soft-X-ray luminosity is very high, and hard-X-ray 
intensity is negligible. There is also a weak power-law 
hard-tail component with an energy spectral slope $\alpha \sim 1.5$. 
In the two component advective flow (TCAF) model (CT95),
the viscous Keplerian disk resides 
in the equatorial plane, while the weakly viscous sub-Keplerian flow 
flanks the Keplerian component both above and below the 
equatorial plane. The two components merge into a single component 
when the Keplerian disk also become sub-Keplerian. It is suggested 
(Chakrabarti, 1990) that close to a black hole, at around $10-15~r_g$, 
($r_g = 2GM_{BH}/c^2$ is the Schwarzschild radius, $M_{BH}$ 
and $c$ are the mass of the black hole and the velocity of
light respectively) the sub-Keplerian flow slows down due  
to the centrifugal barrier and becomes hotter. Chakrabarti 
(1999, hereafter Paper I) shows that this centrifugal 
pressure-supported boundary layer (CENBOL for short) region could be 
responsible for the generation of thermally-driven outflowing
winds and jets and computed the ratio of the outflow to the inflow 
rate assuming a simple conical accretion disk model.

In the present {\it paper}, we compute the {\it absolute} value 
of the outflow rate as a function of the rates of the two inflow 
components, Keplerian and sub-Keplerian. This we do 
analytically following the recently developed  
procedure of obtaining shock locations (Das, Chattopadhyay 
and Chakrabarti, 2001). By dynamically mixing these
two components using solutions of the viscous transonic flows
we obtain the specific energy and angular momentum of the sub-Keplerian
region. We use these pair of parameters to locate shocks in the flow,
compute the compression ratio and from this, the outflow rate. 
We note that as Keplerian matter is increased in the mixture, the shock
compression ratio goes down, and the outflow rate decreases. This is
also the case even from a radiative transfer point of view --  when the 
Keplerian rate is high, the CENBOL region is completely cooled 
and the shock compression ratio $R\sim 1$. Hence in the soft state,
which is due to increase of the Keplerian rate, 
outflow should be negligible.  

In the next Section, we present the governing equations to compute the 
outflow rates using a purely analytical method. We compute
results for both the isothermal and adiabatic outflows. In \S 3, we present our 
results for a single component sub-Keplerian flow.
We also produce examples of realistic disks with Keplerian and 
sub-Keplerian components and obtain outflow rates as functions 
of the inflow parameters.  In \S 4, we discuss our results
and draw conclusions.

\section{Model Equations}

We consider matter accreting on the equatorial plane of a 
Schwarzschild black hole. Spacetime around the black hole is 
described by the Paczy\'nski-Wiita pseudo-Newtonian potential 
$\phi=\frac{GM_{BH}} {r-2GM_{BH}/c^2}$ (Paczy\'nski \& Wiita,
1980) where $M_{BH}$ is the mass of the black hole and $G$, 
$c$ are the gravitational constant and velocity of light respectively.
Here, $r$ is the radial distance from the origin of 
the co-ordinate in which the black hole is treated at the centre.
We use geometric units in which all the length, time and 
velocity scales are measured in units of $2GM_{BH}/c^2$, 
$2GM_{BH}/c^3$ and $c$ respectively. In future, we use  
$r$ to denote non-dimensional distance, $\vel$ and $a$ 
to denote the non-dimensional radial velocity and adiabatic 
speed of sound respectively. In accretion or outflow,
we assume that the viscous stress is negligible so that
matter moves with a constant specific angular momentum.
Indeed, even if viscosity is not negligible, the 
transport of angular momentum is slow compared to the
infall timescale. Hence, matter can have almost constant 
specific angular momentum.

In this case, the radial momentum equation for a non-dissipative 
flow in vertical equilibrium is given by (Chakrabarti, 1989),
$$
\vel\frac{d\vel}{dr}+\frac{1}{\rho}\frac{dP}{dr}
-\frac{\lambda^2}{r^3} +\frac{1}{2(r-1)^2} =0.
\eqno{(1)}
$$
Integrating this, we obtain the conserved specific energy of the flow,
$$
\eng_{v} = \frac{1}{2}\vel^2+na^2 +\frac {\la^2}{2r^2}-\frac{1}{2(r-1)} , 
\eqno{(2)}
$$
where $n$ is the polytropic index of the inflow and $\lambda$
is the specific angular momentum. In Eq. (1), $P$ and $\rho$ are
thermal pressure and  density respectively, $v$ is the infall velocity
and $a=\sqrt(\gamma P/ \rho)$ is the adiabatic sound speed.

The mass flux conservation equation in a flow which 
is in vertical equilibrium is given by,
$$
{\dot M_{in}}= 4\pi\rho \vel r h(r)= \Theta_{in} \rho_s \vel_s r^2_s ,
\eqno{(3)}
$$
where $\Theta_{in}(=\sqrt{\frac {2n}{n+1}}4\pi a_s r^{1/2}_s)$ 
is the solid angle subtended by the inflow at the CENBOL boundary.
Subscripts $s$ denotes the quantities at shock (CENBOL boundary) 
and $h(r)=\sqrt{\frac{2}{\gamma}} ar^{1/2}(r-1)$ 
is the half-thickness of the disk in vertical equilibrium
at a radial distance $r$. 

A sub-Keplerian flow with a positive energy 
will pass through the outer sonic point and depending on whether the 
Rankine-Hugoniot condition is satisfied or not, a standing 
shock may form (Chakrabarti, 1990; Chakrabarti 
1996). If a standing shock forms, then the post-shock flow would become 
hotter and would emit hard X-ray radiation. This CENBOL region
behaves similarly to the boundary of a normal star; it would be expected
to drive outflows. Using Eqs. (2) and (3), it is easy to 
obtain shock locations (i.e., outer surface of the CENBOL) 
analytically. Briefly, the procedure to obtain shocks 
involves the following steps:

\noindent (a) For a given pair of specific energy $\eng_v$ and angular momentum
$\la$, one obtains a quartic equation for the sonic point
and solves it for the three sonic points located outside the horizon.
Two of them are saddle type or `X' type sonic points and one is
a centre type or `O' type sonic point.

\noindent (b) From the inner and the outer `X' type points, Mach
numbers are expressed as polynomials of radial distance $r$. These
Mach number expressions satisfy constraints that they must have
appropriate values at the sonic points.

\noindent (c) In addition, it is enforced that the Mach number invariants at the
shock location are also satisfied ($r_s$).

\noindent (d) The resulting equation becomes quartic in $r_s$ and the shock locations
are obtained from its solution.

Details are discussed in Das et al. (2001). We consider only 
the region of the inflow parameter space ($\eng_v,\ \la$)
that is able to produce standing shocks.

In the pre-shock region, matter is cooler and is sub-Keplerian. 
Assuming $\eng_v\sim 0$ (freely falling condition) and $a\sim 0$ (cool gas)
in presence of angular momentum, matter will fall with a velocity,
$$
\vel(r)=\left[\frac {1}{r-1}-\frac {\la^2}{r^2} \right]^{1/2} .
\eqno{(4)}
$$
Using this, from Eq. (3) the density distribution can be obtained.

At the shock $r=r_s$, i.e., the boundary of the CENBOL, 
the compression ratio is given by,
$$
R=\frac{\Sigma_+}{\Sigma_-}=\frac{h_+(r_s) \rho_{+}(r_{s})}
{h_-(r_s) \rho_{-}(r_{s})} = \frac {\vel_-}{\vel_+},
\eqno{(5)}
$$
where subscripts ``$-$'' and ``$+$'' refer, respectively,
to quantities before and after the shock.  Here, $\Sigma$ 
is the density of matter integrated vertically 
$(\Sigma\sim \rho h)$ and the second `$=$' sign was written using 
the mass flux conservation equation given above (Eq. 3).

At the shock, the total pressure (thermal and ram pressure) is balanced:
$$
W_{-}(r_{s})+\Sigma_{-}(r_{s})\vel^2_{-}(r_{s})=W_{+}(r_{s})+
\Sigma_{+}(r_{s})\vel^2_{+}(r_{s}) , 
\eqno{(6)}
$$
where $W$ is the pressure of the gas integrated vertically.

We assume that in the pre-shock region, the thermal pressure 
is small in comparison to the ram pressure, 
$$
W_{+}(r_{s})=\frac {R-1}{R}\Sigma_{-}(r_{s})\vel^2_{-}(r_{s}).
\eqno{(7)}
$$

The isothermal sound speed in the post-shock region is obtained from:
$$
C^2_{s}=\frac {W_{+}}{\Sigma_{+}}=\frac {R-1}{R^2}\vel^2_{-}
=\frac {1}{f_{0}}\left[\frac {r^2_{s}-\la^2(r_{s}-1)} 
{r^2_{s}(r_{s}-1)}\right],
\eqno{(8)}
$$
where, $f_{0}=\frac {R^2}{R-1}$.

Up to the sonic point matter moves slowly and the density is higher.
Since the outflow would take place in a sea of radiation,
the momentum deposition is likely to be effective.
With the electron number density $n_e \propto r^{-3/2}$,
yet photon number density $n_\gamma \propto r^{-2}$, it is 
easier to deposit momentum only close to the black hole.
In radiation driven outflows from the stellar surface, it is 
customary to assume flows to be isothermal until the sonic point.
We first compute outflow rates making this assumption.
Later we drop this assumption and show that the 
general behaviour remains similar.
In addition, we assume that there is very little
rotation in the outflow. There is no a priori
reason to assume this, except that there is no observational support
of rotation in the jet and it is possible that due to radiative
viscosity most of the angular momentum is transported
very close to the black hole. Furthermore, it has been 
observed that the effect of angular momentum in the outflow
is to bring the sonic points closer to the black hole,
especially away from the axis (Sakurai, 1985; Chakrabarti, 1986).
The general effect would produce a transverse structure in the
jet which we ignore in the present solution. It was shown
(Das \& Chakrabarti, 1999) that in presence of angular motion the
conical outflow is to be replaced by an annular flow confined by the 
centrifugal barrier and the funnel wall. Generally speaking, the 
outflow surface varies as $r^{3/2}$. However, the inflow surface area
is still proportional to $r^2$. Because of this asymmetry, the problem is
no longer tractable analytically and is beyond the scope of the present
paper.

\subsection{When the outflow is isothermal}

The radial momentum balance equation in the outflow is given by
$$
\vel \frac {d\vel}{dr}+\frac {1}{\rho}\frac {dP}{dr}+
\frac {1}{2(r-1)^2}=0 ,
\eqno{(9)}
$$
and the continuity equation is given by
$$
\frac {1}{r^2}\frac {d}{dr}(\rho \vel r^2)=0.
\eqno{(10)}
$$
Eliminating $\frac {d\rho}{dr}$ from above two equation we get
$$
\frac {d\vel}{dr}=\frac {N}{D},
\eqno{(11)}
$$
where $N=\frac {2C^2_{s}}{r}-\frac {1}{2(r-1)^2}$ 
and $D=\vel-\frac {C^2_{s}}{\vel}$.

To obtain the sonic point condition, we put $N=0$ and $D=0$ and get,
$\vel(r_c)=C_{s}$, and $r_c=\frac {1+8 C^2_s \pm \sqrt{1+16 C^2_s}}{8 C^2_s}$,
where the subscript $c$ denotes the quantities at the sonic point in the outflow.

Integrating the radial momentum equation, considering the sonic 
point condition, we have,
$$
C^2_sln \rho_+ - \frac {1}{2(r_s-1)}=
\frac {1}{2}C^2_s + C^2_sln\rho_c-\frac {1}{2(r_c-1)}.
\eqno{(12)}
$$ 
Here, we have ignored the radial velocity in the outflow 
at the boundary of the shock. Using the notations 
$\rho(r_c)=\rho_c$ and $\rho(r_s)=\rho_+$, we obtain,

$$
\rho_c=\rho_+exp[-f],
\eqno{(13)}
$$
where $f=\frac {1}{2}- \frac {1}{2C^2_s}\frac {r_s-r_c} {(r_s-1)(r_c-1)}$.

The outflow rate is given by
$$
\dot M_{out}=\Theta_{out} \rho_c \vel_c r^2_c ,
\eqno{(14)}
$$
where $\Theta_{out}$ is the solid angle subtended by the outflow.

From Eq.(3) \& Eq. (14) we get, 
$$
\frac {\dot M_{out}}{\dot M_{in}}=R_{\dot m}=\frac 
{\Theta_{out}}{\Theta_{in}}\left[\frac {r^2_s(r_s-1)}
{r^2_s-\la^2(r_s-1)} \right]^{-{1/2}}\frac {RC_s r^2_c} {r_s(r_s-1)}exp[-f] .
\eqno{(15)}
$$
The above relation is very similar to that obtained in 
Paper I when the effects of rotation in the inflow were ignored. However, 
there the ratio $R_{\dot m}$ was a function of $R$ alone. 
In the present analysis, $R$ is computed self-consistently from the specific
energy and the specific angular momentum of the flow:
$$
R = \frac{\Sigma_+}{\Sigma_-} = \frac{\vel_-}{\vel_+}=
\left [ \frac {\frac {1}{2} M_+^2 + n}{\frac {1}{2} M_-^2 +n} 
\right ]^{1/2} ,
\eqno{(16)}
$$
where pre-shock and post-shock Mach numbers $M_- ({\cal E}, \la)$ and 
$M_+({\cal E}, \la)$ are computed analytically from Das et al. (2001). 

\subsection{When the outflow is adiabatic}

At the other extreme, when the energy of the outflow does not change,
one can also obtain an analytical expression for the outflow rate
assuming the $r_s >>\lambda^2$. In this case, the entropy density of the
flow in the post-shock region is the same at the entropy density 
of the entire outflow and the specific energy is also conserved
along the outflow. We assume that the turbulence generated at the
CENBOL has effectively transported angular momentum away. Thus, the
energy conservation equation gives
$$
na_s^2 -\frac{1}{2r_s} = \frac{2n+1}{2}a_c^2 -\frac{1}{2r_c},
\eqno {(17)}
$$
where the left hand side is the energy at the CENBOL ($r=r_s$) and the right
hand side is at the sonic point ($r=r_c$) of the outflow where $u_c=a_c$ 
has been used. $n=\frac{1}{\gamma-1}$ is the polytropic constant. 
In a Bondi (in or out)flow, $a_s^2=1/4r_c$. At the
CENBOL, $a_s^2 = \gamma C_s^2$, where $C_s$ is the isothermal sound speed
(Eq. 8). Using these, one obtains (assuming $r_s>>\lambda^2$)
$$
\frac{r_c}{r_s}=\frac{2n-3}{4(\frac{2n\gamma}{f_0}-1)} ,
\eqno{(18a)}
$$
and
$$
\frac{a_c^2}{a_s^2}=\frac{f_0 r_s}{4\gamma r_c} .
\eqno{(18b)}
$$
In an adiabatic flow with an equation of state $P=K\rho^\gamma$
(where $K$ is a constant and a measure of entropy), one obtains,
assuming, $K_c=K_s$ ,
$$
\frac{\rho_c}{\rho_s} = \left [\frac{a_c^2}{a_s^2}\right]^n .
\eqno{(19)}
$$
From these relations one obtains the ratio of the outflow
to the inflow rate as
$$
R_{\dot m}= \frac{\Theta_o}{\Theta_i} (\frac{f_0}{4\gamma})^3 \frac{R}{2}
\{\frac{4}{3}[\frac{8(R-1)}{R^2} -1]\}^{3/2}  .
\eqno{(20)}
$$
Here, we have used $n=3$ for a relativistic flow. The nature of this function
will be discussed below.

\section{Outflow rates from inflow parameters}

In Eq. (15), we presented the outflow/inflow rate ratio 
as a function of the compression ratio of the flow at the 
shock. The compression ratio is obtained from the specific energy
and angular momentum using Eq. (16). First, we employ analytical
means to obtain this for a single component sub-Keplerian disk. 
Second, we use a two component Keplerian/sub-Keplerian disk 
to actually compute these parameters from more fundamental 
parameters such as accretion rates and viscosity.

\subsection{Single component sub-Keplerian Flows}


\begin {figure}
\resizebox{\hsize}{!}{\includegraphics{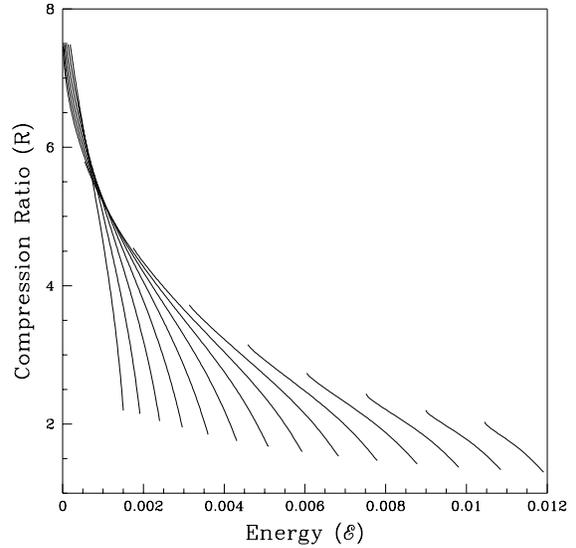}}
\caption{
Variation of the compression ratio of the
shocks as a function of specific energy ${\cal E}$ 
and angular momentum $\lambda$ as obtained from the analytical 
solution. $\lambda$ varies from $1.57$ (right) to $1.79$ (left). 
Curves are drawn at intervals of $d\lambda=0.02$.}
\end{figure}

In Fig. 1, we  plot the analytical solution
of the compression ratio $R$ as a function of the flow
parameters: specific energy ${\cal E}$ and the specific 
angular momentum $\lambda$. The shock strength generally 
increases when energy decreases and the angular momentum 
increases. This is because for low energy, the outer sonic
point and the shock form
very far away and the Mach number jumps from a very large number 
to a very small number. If the angular momentum is decreased,
shock is  produced only if the specific energy is high,
i.e., if the sonic points and the shocks are very close to the black hole.
Here, flow becomes subsonic before its Mach number could be very high.

\begin {figure}
\resizebox{\hsize}{!}{\includegraphics{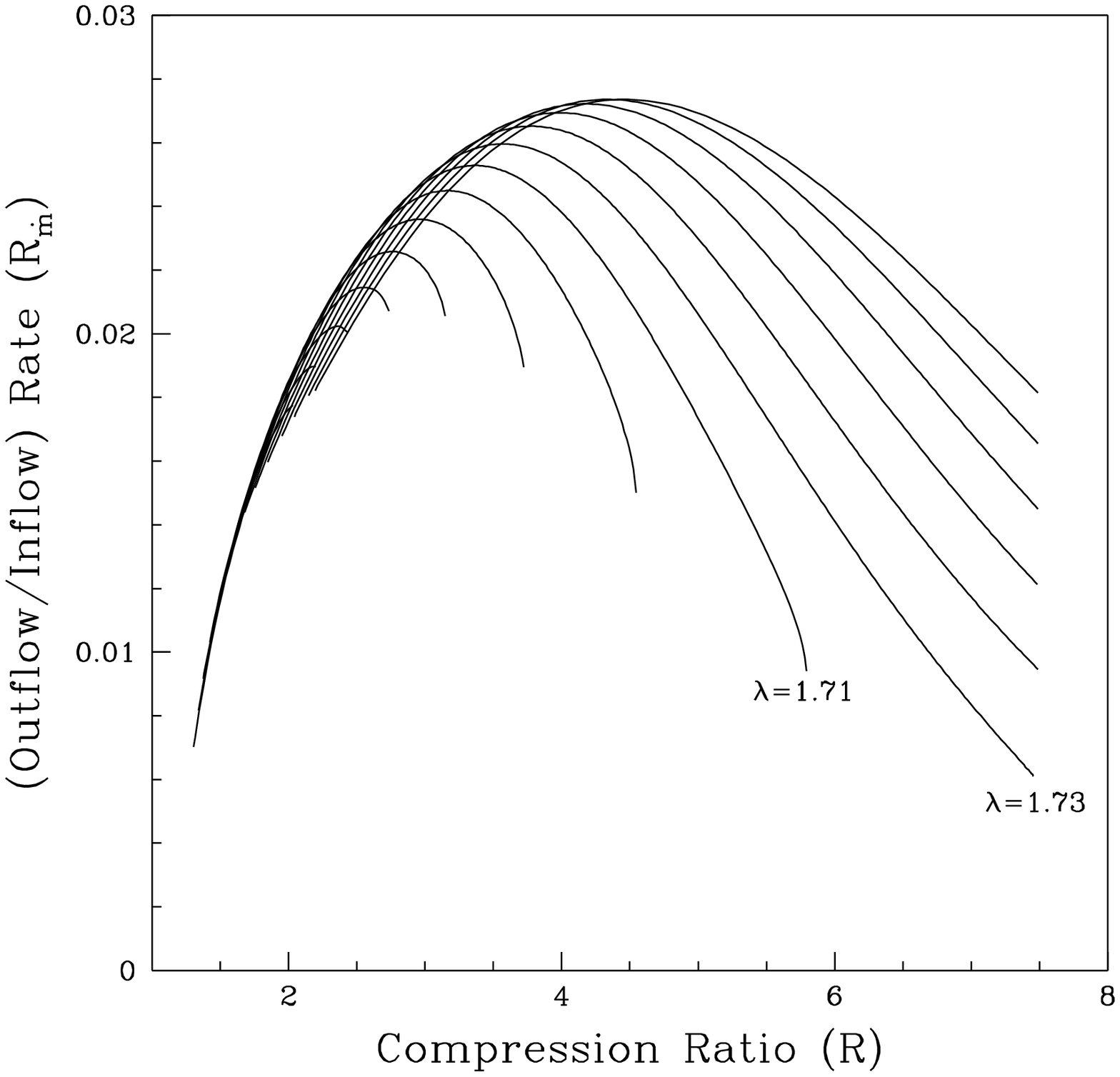}}
\noindent {{\bf Fig. 2a.} Variation of the ratio of outflow to inflow 
rates $R_{\dot m}$ as a function of compression ratio 
for various specific angular momenta. $\lambda=1.57$ (inner 
most) to $1.83$ (outer most). Curves are drawn at intervals 
of $d\lambda=0.02$. Outflow rate is maximum at some intermediate 
compression ratio. }
\end{figure}

Figure 2a shows the principle result of our work when only one 
sub-Keplerian accretion is chosen as the inflow. We plot the 
ratio $R_{\dot m}$ 
for a large number of specific angular momenta of the flow ranging from 
$1.57$ (innermost) to $1.83$ (outermost) at intervals of 
$d\lambda=0.02$. The curves are drawn for all ranges of 
specific energy ${\cal E}$ for which shocks are formed. 
Along the X-axis the compression ratio $R$ of these shocks is written. 
Here to compute solid angles of the inflow and the outflow, we 
assume the half opening angle of the outflow to be 10$^o$. Therefore,
$\Theta_{out}=\pi^3/162$. $\Theta_{in}$ is given in the discussion following
Eq. (3). In Paper I, the compression ratio $R$ was assumed to be a 
parameter and 
no angular momentum was assumed a priori. Presently, 
we show the dependence on angular momentum. 
The general character, namely, 
that the outflow rate is negligible when the shock 
is weak ($R \sim 1$) and falls off gradually 
for strongest shock ($R\rightarrow 7$), remains the 
same as in Paper I, however. There is a peak at about $R_{\dot m}\sim2.8 \%$.
Note that for a given $R$, $R_{\dot m}$ 
increases monotonically with specific angular 
momentum $\lambda$. This is because density of 
CENBOL rises with $\lambda$. The curves 
corresponding to $\lambda=1.71$ and $1.73$ 
are specially marked since there is a clear
difference in tendency of the variation of $R_{\dot m}$. 
For instance, below $\lambda\sim 1.72$, very strong
shocks are not possible at all and as a result the 
outflow rate has a lower limit. For $\lambda\gsim 1.72$ 
such a limit does not exist. 

The general behaviour of the outflow rate can be understood in
the following way: when
shocks are strong, they form very far out, and thus, even though
the CENBOL area (which is basically the area of the base
of the jet) increases, the net outflow rate is low. When the 
shock forms very close to the black hole, the temperature is
high, and thus the outflow velocity is larger, but the CENBOL 
surface area goes down. Thus the product is low. For the
intermediate cases the net effect is larger.

\begin {figure}
\resizebox{\hsize}{!}{\includegraphics{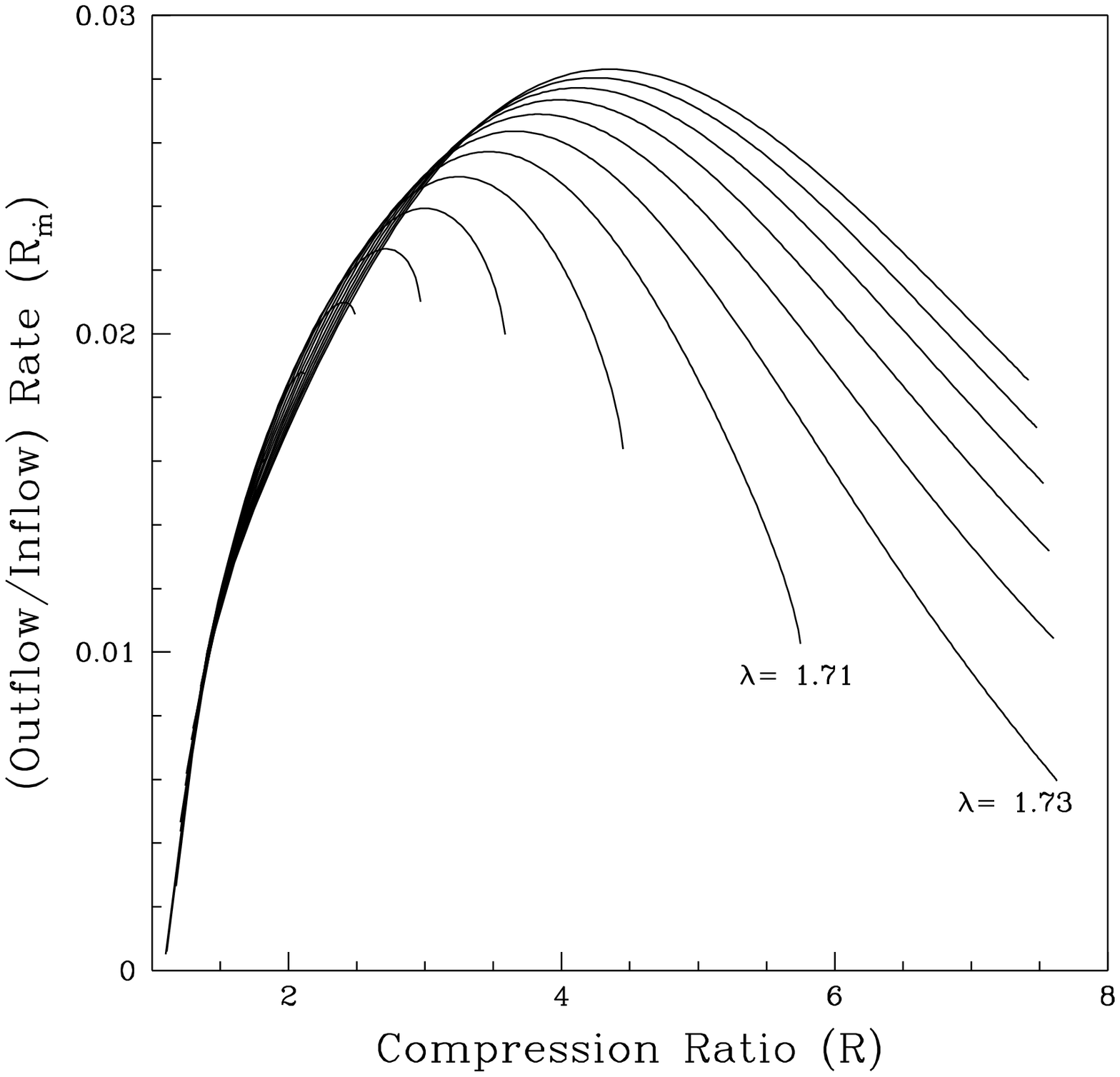}}
\noindent {{\bf Fig. 2b:}  Same as Fig. 2a except that curves are
drawn for the exact numerical solution.}
\end{figure}

For comparison with the analytical work presented 
in Fig. 2a, in Fig. 2b we present a similar diagram 
drawn using a numerical computation of the shock
locations (Chakrabarti, 1989). Excellent agreement
between these two figures
implies that the approximations on which the 
analytical work was based are justified. 
All the features are reproduced well in Fig. 2a, 
except that for the weakest shocks outflow
rate is not as low as in the numerical 
calculation of Fig. 2b.    

\begin {figure}
\resizebox{\hsize}{!}{\includegraphics{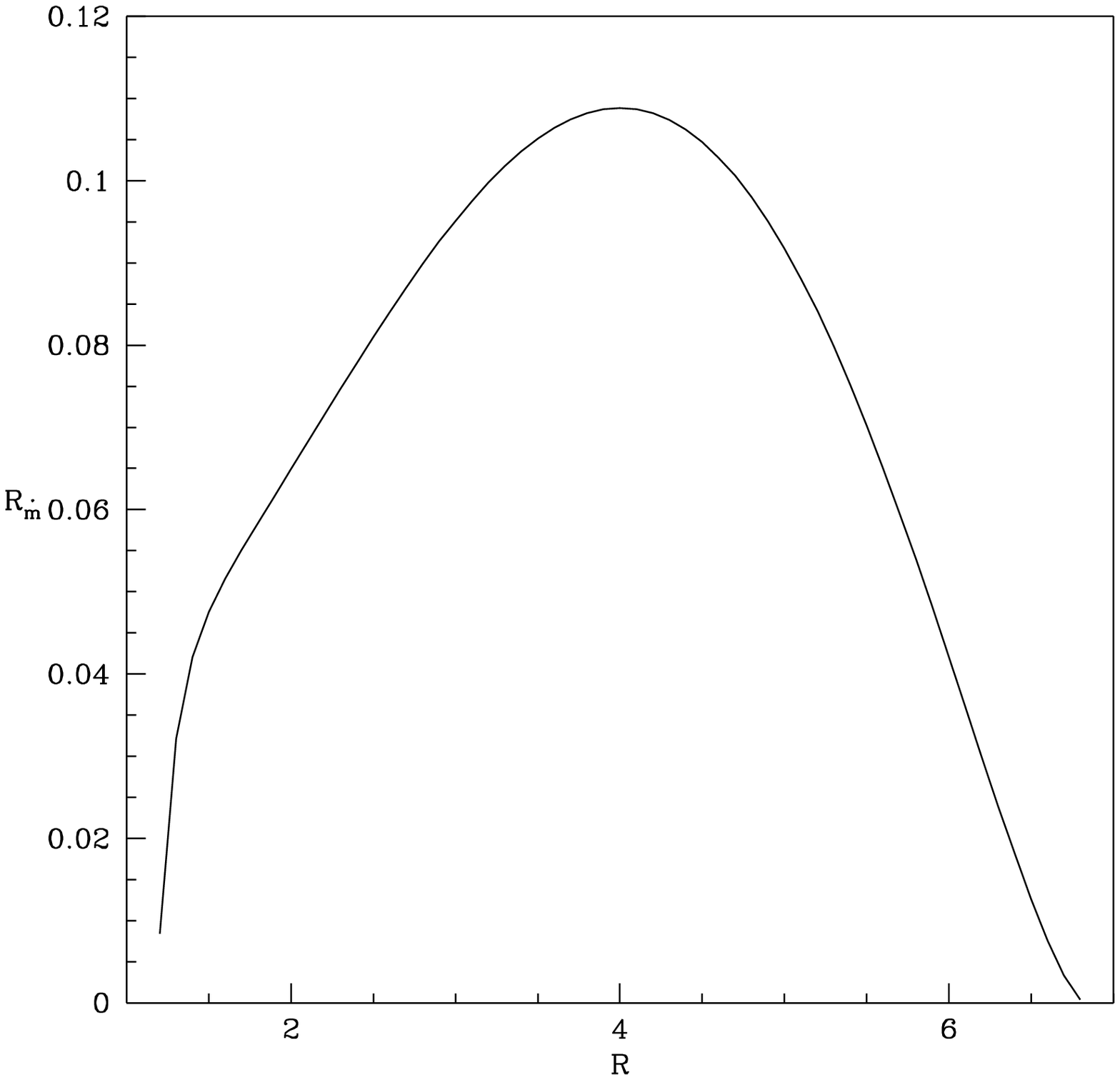}}
\noindent {{\bf Fig. 3:} Ratio of the Outflow and the Inflow rates
as a function of the compression ratio of the inflow when the outflow
is adiabatic. The general nature of the function remains the same
as that of the isothermal outflow.}
\end{figure}

We now present the nature of $R_{\dot m}$ when the outflow is also chosen 
to be adiabatic in Fig. 3. We used $\Theta_o/\Theta_i \sim 0.1$ for reference. 
We observe that the peak is still located at around $R= \sim 4$ and the outflow
rate drops for very strong ($R\sim 7$) and very weak ($R \sim 1$) shocks.
We therefore believe that our conclusion about the behaviour of 
$R_{\dot m}$ is generic.

\subsection{Two component advective flows}

Chakrabarti \& Titarchuk (1995) proposed
that the spectral properties are better 
understood if the disk solutions 
of sub-Keplerian flows are included along with the
Keplerian flows. Recently, Smith, Heindl and Swank (2001), Smith et al. (2001),
Miller et al. (2001) found
conclusive evidence of these two components in many of 
the black hole candidate accretion flows. 
While the matter with higher viscosity flows
along the equatorial plane as a Keplerian disk (of 
rate ${\dot M}_K$), sub-Keplerian halo matter (of rate 
${\dot M}_h$) with lower viscosity flanks the Keplerian 
disk above and below (Fig. 3a). Since the inner boundary condition
on the horizon forces the flow to be sub-Keplerian, 
irrespective of their origin (Chakrabarti, 1990, 1996) 
matter mixes (at say, $r=r_{tr}$) from both the Keplerian 
and sub-Keplerian flows before entering a black hole
to form a single component sub-Keplerian with an average 
energy and angular momentum of ${\cal E}$ and ${\lambda}$
respectively. The specific energy and angular 
momentum of the mixed flow is computed from:
$$
{\cal E}= \frac{{\dot M}_K {\cal E}_K +{\dot M}_h {\cal E}_h}
{{\dot M}_K+{\dot M}_h},
\eqno{(21)}
$$
and
$$
\lambda= \frac{{\dot M}_K \lambda_K +{\dot M}_h \lambda_h}
{{\dot M}_K+{\dot M}_h}.
\eqno{(22)}
$$
Here, ${\cal E}_K$, ${\cal E}_h$, $\lambda_K$ and $\lambda_h$ 
are the specific energies and specific angular momentum
of the Keplerian and the sub-Keplerian components at $r=r_{tr}$
respectively. 

\begin {figure}
\resizebox{\hsize}{!}{\includegraphics{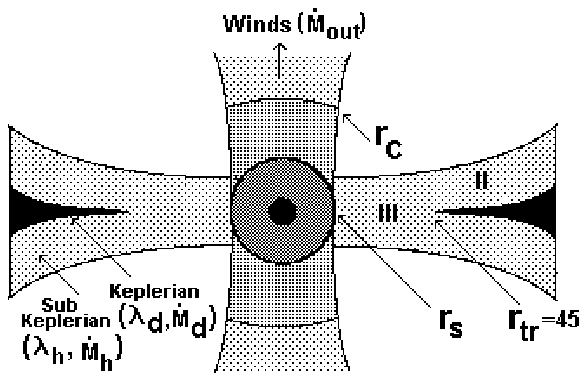}}
\noindent {{\bf Fig. 4a:} Schematic diagram of the cross section of 
two-component accretion flow. See text for details.}
\end{figure}

\begin {figure}
\resizebox{\hsize}{!}{\includegraphics{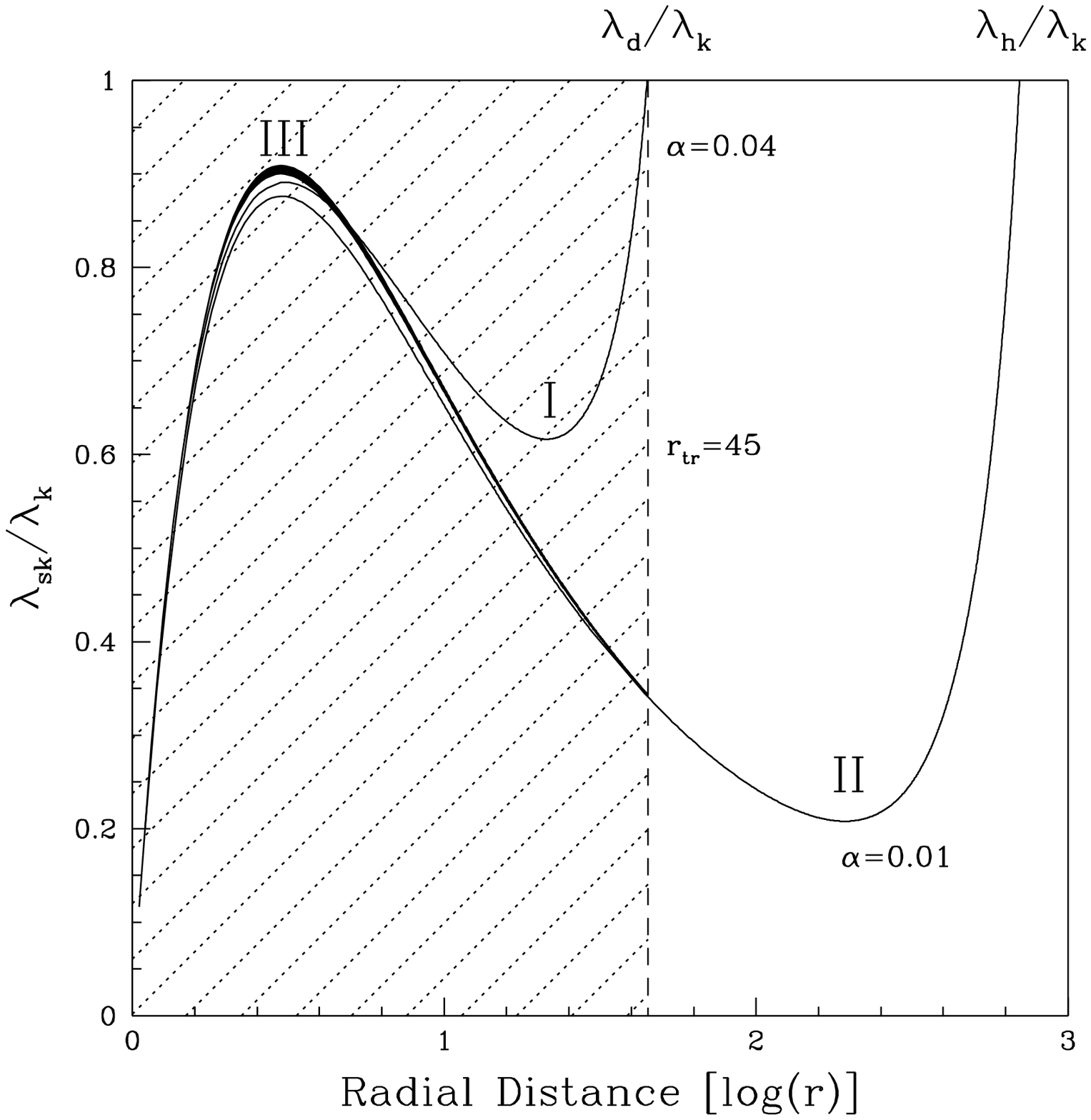}}
\noindent {{\bf Fig. 4b:} Solution of the two-component flow equations for 
two different viscosities. They are merged to form a single
solution as depicted in Fig. 4a. }
\end{figure}

Figure 4a shows a schematic diagram of the cross-section 
of a two-component accretion flow. The transition radius 
($r=r_{tr}$) where the Keplerian disk becomes sub-Keplerian,
and the shock location $r=r_s$, are indicated. Fig. 4b shows 
two solutions (marked I and II) of the equations 
governing a two-component flow 
(Chakrabarti, 1996) where $\lambda_d/\lambda_K$ (Sub-Keplerian 
matter from the Keplerian disk) and $\lambda_h/\lambda_K$ 
(Sub-Keplerian halo) are plotted as a function of the logarithmic 
radial distance. Viscosities chosen for these two components
are $\alpha=0.04$ and $\alpha=0.01$ respectively. 
For $r<r_{tr}=45$ (lightly shaded region) the two sub-Keplerian 
flows mix to create a single component. For simplicity,
we assume viscosity to be negligible in this region. 
Thus, the specific angular momentum and specific energy
computed at $r=r_{tr}$ from Eqs. (21 \& 22) remain constant ($\lambda$)
for $r<r_{tr}$. Dark solid curve (marked III) shows the angular momentum 
distribution $\lambda/\lambda_K$ of all possible  mixtures
of the two components which allow shock formation. We chose a case 
where ${\dot M}_d+{\dot M}_h =2.0 {\dot M}_{Edd}$ and 
vary the Keplerian component ${\dot M_d}$ where ${\dot M}_{Edd}$
is the Eddington accretion rate.

\begin {figure}
\resizebox{\hsize}{!}{\includegraphics{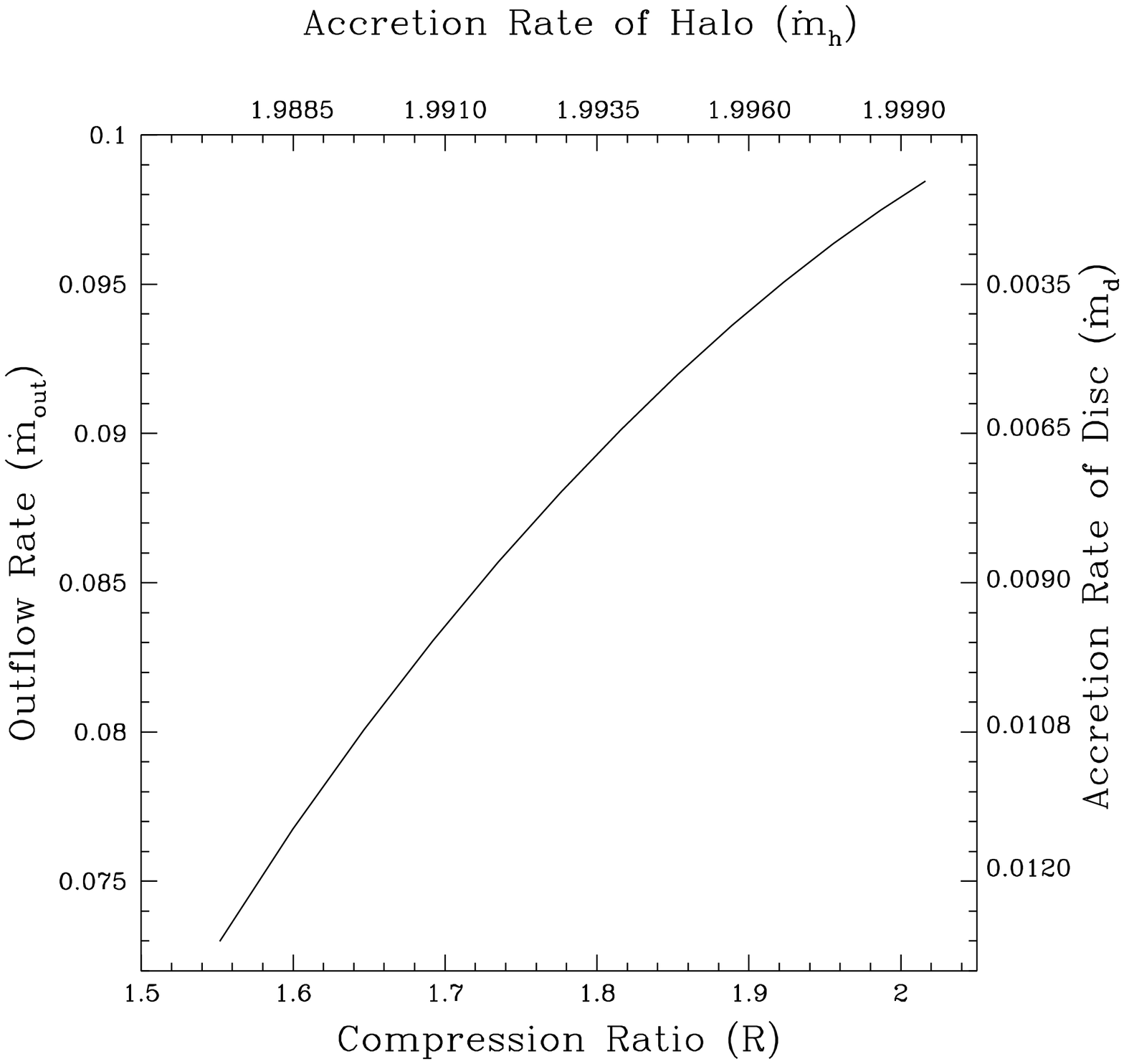}}
\noindent {{\bf Fig. 5:} Variation of outflow rates (left axis) with 
compression ratio at shocks (lower axis). The upper axis
gives the variation of sub-Keplerian accretion rate and 
right axis gives the same for Keplerian accretion rate.
}
\end{figure}

In Fig. 5, the computed outflow rates are shown when
the half opening angle of the outflow is $10^o$. In this case,
$\frac{\Theta_{out}}{\Theta_{in}} \sim \sqrt{\frac{n+1}{2n}}\frac {\pi^2}
{648 a_s r^{1/2}_s}$. The left axis shows the rate of 
outflow ${\dot m}_{out}= {\dot M}_{out}/{\dot M}_{Edd}$ as a function 
of the Keplerian disk rate (right panel) (${\dot m}_d 
={\dot M}_d/{\dot M}_{Edd}$) and the halo rate (upper panel) 
(${\dot m}_h ={\dot M}_h/{\dot M}_{Edd}$). The lower 
axis gives the compression ratio at the shock. The most important 
conclusion that can be drawn here is that the outflow 
rate steadily goes up as the Keplerian disk rate ${\dot m}_d$ 
decreases and the spectrum goes to a harder state. When the 
Keplerian rate is higher, the compression ratio is lower and 
the outflow rate is also lower. This conclusion, drawn completely 
from dynamical considerations, is also found to be true 
from the spectral studies (CT95) 
where it was shown that the post-shock region cools down and 
the shock disappears ($R \rightarrow 1$).  Our work therefore
hints that the outflow would be negligible in softer states.

\section{Discussion and Concluding Remarks}

CT95  pointed out that 
the centrifugal pressure-supported boundary layer (CENBOL) 
of a black hole accretion flow is
responsible for the spectral properties of a black hole candidate.
In this {\it Paper}, we present analytical results to show that
this CENBOL is also responsible for the 
production of the outflows,  and the outflow rate is strongly 
dependent on the inflow parameters, such as specific energy 
and angular momentum.  We showed that in general, the outflow 
rate is negligible when the shock is absent and very small 
when the shock is very strong.  In intermediate strength,
the outflow rate is the highest. As the specific angular 
momentum is increased, the outflow rate is also increased. 
This conclusion is valid when the flow is either isothermal 
or adiabatic.

We also demonstrated how a realistic two-component flow  (TCAF)
consisting of Keplerian and sub-Keplerian components produces a
significant amount of outflow. Since matter close to a black hole
is sub-Keplerian by nature, the two components must mix to form
a single sub-Keplerian flow which has positive specific energy
and almost constant specific angular momentum. We showed that 
as the Keplerian rate of the disk is increased, the outflow rate is 
decreased as the shock compression ratio approaches unity. 
This conclusion, drawn from a dynamical point of view, is 
also corroborated by the spectral behavior as well --- as 
the Keplerian rate is raised, the post-shock region is cooled 
due to inverse Comptonization and the shock disappears. This
reduces the thermal pressure drive and the resulting  outflow rate
is reduced.  

\begin{acknowledgements}
This work is partly supported by a project (Grant No. SP/S2/K-14/98)
funded by Department of Science and Technology (DST), Govt. of India.
\end{acknowledgements}

{}
\end{document}